\documentclass[12pt,reqno]{amsart}

\usepackage{amssymb,amsthm,amsmath,mathrsfs}
\usepackage{enumerate}
\usepackage{fullpage}
\usepackage{graphicx}
\usepackage{float}
\usepackage{subfig}

\newtheorem{theorem}{Theorem}[section]
\newtheorem{proposition}[theorem]{Proposition}

\def\Rnum{{\mathbb R}}
\def\X{\mathrm{X}}
\def\pr{\mathrm{pr}}

\def\Hop{{\mathcal H}}
\def\intx{\partial_x^{-1}}

\def\Ref#1{Ref.\cite{#1}}

\DeclareMathOperator{\sech}{sech}

\begin{document}

\title{Conservation laws, symmetries, and\\ line solitons of a Kawahara-KP equation}

\author{
Almudena P. M\'arquez$^{1\dagger}$,
Mar\'ia L. Gandarias$^1$,
Stephen C. Anco$^2$
\\\\
${}^{1}$D\lowercase{\scshape{epartment}} \lowercase{\scshape{of}} M\lowercase{\scshape{athematics}}\\
U\lowercase{\scshape{niversity of}} C\lowercase{\scshape{adiz}}\\
11510 P\lowercase{\scshape{uerto}} R\lowercase{\scshape{eal}}, C\lowercase{\scshape{adiz}}, S\lowercase{\scshape{pain}}\\
\\
${}^2$D\lowercase{\scshape{epartment}} \lowercase{\scshape{of}} M\lowercase{\scshape{athematics and}} S\lowercase{\scshape{tatistics}}\\
B\lowercase{\scshape{rock}} U\lowercase{\scshape{niversity}}\\
S\lowercase{\scshape{t.}} C\lowercase{\scshape{atharines}}, ON L2S3A1, C\lowercase{\scshape{anada}} \\
}

\thanks{${}^\dagger$almudena.marquez@uca.es}


\begin{abstract}
A generalization of the KP equation involving higher-order dispersion is studied. 
This equation appears in several physical applications.
As new results, 
the Lie point symmetries are obtained and used to derive conservation laws
via Noether's theorem by introduction of a potential which gives a Lagrangian formulation for the equation. 
The meaning and properties of the symmetries and the conserved quantities 
are described. 
Explicit sech-type line wave solutions are found 
and their features are discussed. 
They are shown to describe dark solitary waves on a background 
which depends on a dispersion ratio and on the speed and direction of the waves. 
The zero-background case is explored. 
\end{abstract}

\maketitle

\section{Introduction}

Dispersive wave equations arise in many areas of applied mathematics and physics.
One prominent example is the Korteweg-de Vries (KdV) equation
$u_t + \alpha u u_x+ \beta u_{xxx} =0$ ($\alpha,\beta\neq0$)
which models shallow water waves and is a bi-Hamiltonian integrable system.
The dispersion is caused by the $u_{xxx}$ term,
while the other term $uu_x$ describes weakly nonlinear advection.

Higher-order dispersive effects are relevant in certain physical applications, 
such as capillary-gravity water waves \cite{Has}
and magneto-acoustic waves in plasmas \cite{Kak,KakOno}. 
The addition of a fifth-order term $u_{xxxxx}$ to the KdV equation yields 
the Kawahara equation
$u_t + \alpha u u_x +\beta u_{xxx} + \gamma u_{xxxxx} =0$ 
($\alpha,\beta,\gamma\neq0$). 
It has a Hamiltonian structure, but in contrast to the KdV equation,
it is not an integrable system.
Numerical studies of solutions \cite{Kaw} have shown that 
the higher-order dispersion can produce both oscillatory and monotone 
solitary wave solutions. 

In two spatial dimensions, 
wave phenomena that exhibit weak transversality and weak nonlinearity 
are modelled by the Kadomtsev–Petviashvili (KP) equation \cite{KadPet}
$u_t + \alpha u u_x +\beta u_{xxx} = \sigma \partial_x^{-1} u_{yy}$
($\alpha,\beta,\sigma\neq0$). 
Like the KdV equation, 
it is an integrable system with a bi-Hamiltonian structure 
and it possesses line soliton solutions as well as lump soliton solutions in the case $\beta/\sigma>0$. 
(See e.g.\ \Ref{AblCla-book}).

Inclusion of higher-order dispersion leads to a fifth-order generalization of the KP equation \cite{Kar1993}
\begin{equation}\label{KKPeqn}
u_t + \alpha u u_x +\beta u_{xxx}+\gamma u_{xxxxx}  =\sigma \partial_x^{-1} u_{yy}
\end{equation}
which we will call the Kawahara-Kadomtsev-Petviashvili equation, 
or ``K-KP'' equation in short form. 
One main physical application is modelling long water waves 
in a shallow water regime with strong dispersion, 
where $u$, after suitable dimensionful scaling, 
describes both the wave speed and the wave amplitude
\cite{Har,HunSch,IliSem}.
Another physical application is modelling plasma waves with strong dispersion \cite{Kaw}. 

Equation \eqref{KKPeqn} has a Hamiltonian structure
\begin{equation}\label{hamileqn}
u_t = D_x(\delta H/\delta u)
\end{equation}
where $D_x$ is a Hamiltonian operator \cite{Olv-book},
and where the Hamiltonian is given by 
\begin{equation}\label{hamil}
H = \int_{\Omega} ( -\tfrac{1}{6}\alpha u^3 +\tfrac{1}{2} \beta u_x{}^2 -\tfrac{1}{2}\gamma u_{xx}{}^2 + \tfrac{1}{2} \sigma (\partial_x^{-1} u_y)^2 )\,dx\,dy
\end{equation}
on any fixed spatial domain $\Omega\subseteq\Rnum^2$. 
Here $\delta/\delta u$ is the variational derivative. 
There has been work on numerical solutions in \Ref{KarBel,AbrSte}, 
the Cauchy problem in \Ref{SauTzv1999,SauTzv2000,CheLiMia},
and transverse stability of solitary waves that move in the $x$-direction in \Ref{Kar1993,EsfLev}. 

Two open aspects about the K-KP equation to-date 
are its symmetries and conservation laws,
and its exact solitary line wave solutions. 
In the present work, 
we provide a complete classification of all point symmetries
from which we obtain conservation laws by means of Noether's theorem
using a Lagrangian formulation in terms of a potential. 
We describe the meaning and main features of 
the symmetries and the conserved quantities, 
and we also explain how the Hamiltonian structure provides 
a direct connection between the conservation laws and the symmetries.
In addition, we obtain explicit sech-type line wave solutions
that move transversely with respect to the $x$-direction, 
and we show that they describe dark solitary waves on a background 
whose size depends on the dispersion ratio and on the speed and direction of the waves.
We explore the zero-background case 
and discuss the main kinematic features of the solutions. 
Finally, we compare these solutions to the explicit sech-type solitary waves 
known \cite{Alb,Sir,Nat} for the Kawahara equation.

Note that, by scaling transformations on the variables $x$, $y$, $t$, and $u$,  
the coefficients in equation \eqref{KKPeqn} can be set to 
$\alpha =1$, $\gamma=1$, $\sigma^2 = 1$. 
For the sequel, we consider this scaled form of the equation:
\begin{equation}\label{KKPeqn.scal}
u_t + u u_x +\beta u_{xxx}+ u_{xxxxx} =\sigma \partial_x^{-1} u_{yy}, 
\quad
\beta\neq 0,
\quad
\sigma=\pm 1.
\end{equation}
Through introduction of a potential 
\begin{equation}\label{pot}
u=v_x , 
\end{equation}
equation \eqref{KKPeqn.scal} becomes a local PDE
\begin{equation}\label{KKPpoteqn}
v_{tx} + v_x v_{xx} + \beta v_{xxxx} + v_{xxxxxx} = \sigma v_{yy}
\end{equation}
which is useful as it will avoid nonlocalities in the formulation of 
symmetries and conservation laws. 

We will refer to the cases 
$\sigma =1$ as K-KP I and $\sigma=-1$ as K-KP II, respectively. 
This is motivated in analogy with the usual terminology for the KP equation,
which distinguishes the two cases for the sign of the ratio of 
the highest derivative terms in $x$ and $y$.

\section{Point symmetries}

Lie point symmetries are a basic structure 
as they can be used to find invariant solutions
and yield transformations that map the set of solutions $v(x,y,t)$ into itself.
A general discussion of symmetries and their applications to PDEs 
can be found in \Ref{Olv-book,BCA-book,Anc-review}.

An infinitesimal point symmetry of the potential K-KP equation \eqref{KKPpoteqn}
\begin{equation}
\X =\xi^x(x,y,t,v)\partial_x + \xi^y(x,y,t,v)\partial_y +\tau(x,y,t,v)\partial_t+\eta(x,y,t,v)\partial_v
\end{equation}
is a generator of a point transformation 
whose prolongation leaves the equation invariant. 
Every point symmetry generator can be expressed in an equivalent, characteristic form
\begin{equation}
\hat\X =P\partial_v,
\quad
P=\eta(x,y,t,v) -\xi^x(x,y,t,v)v_x -\xi^y(x,y,t,v)v_y -\tau(x,y,t,v)v_t
\end{equation}
which acts only on $v$.
Invariance of the potential K-KP equation \eqref{KKPpoteqn} is given by the condition
\begin{equation}\label{symm-deteqn}
0=D_t D_x P +v_x D_x^2 P + \beta D_x^4 P + D_x^6 -\sigma D_y^2 P
\end{equation}
holding for all solutions $v(x,y,t)$ of equation \eqref{KKPpoteqn}.
This condition \eqref{symm-deteqn} splits with respect to derivatives of $v$,
which yields an overdetermined system of equations
on $\eta(x,y,t,v)$, $\xi^x(x,y,t,v)$, $\xi^y(x,y,t,v)$, $\tau(x,y,t,v)$, 
along with $\beta\neq0$ and $\sigma=\pm 1$. 
It is straightforward to set up and solve this determining system by Maple 
using the `rifsimp' command. 

We obtain the following result.

\begin{theorem}
The Lie point symmetries admitted by the K-KP potential equation \eqref{KKPpoteqn}
are generated by
\begin{align}
& 
\X_1=\partial_t,
\label{t-transl}\\
&
\X_2 =f(t) \partial_x  + ( x f'(t) + \tfrac{1}{2\sigma} y^2 f''(t) ) \partial_v, 
\label{gen.x-transl}\\
&
\X_3 = \tfrac{1}{2\sigma} y f'(t) \partial_x+ f(t) \partial_y  + \tfrac{1}{2\sigma} y ( x f''(t) + \tfrac{1}{6\sigma}y^2 f'''(t) ) \partial_v, 
\label{gen.y-transl}\\
&
\X_4=f(t) \partial_v,  
\label{shift1}\\
&
\X_5 = y f(t) \partial_v, 
\label{shift2}
\end{align}
where $f(t)$ is an arbitrary function. 
\end{theorem}

These symmetry generators can be prolonged to act on $v_x=u$ by 
\begin{equation}
\pr^{(1)}\X = \X + \eta^u\partial_u := \X^u, 
\quad
\eta^u = D_x \eta
\end{equation}
which yields 
\begin{align}
& 
\X^u_1=\partial_t,
\label{u.t-transl}\\
&
\X^u_2 =f(t) \partial_x + f'(t) \partial_u, 
\label{u.gen.x-transl}\\
&
\X^u_3 = \tfrac{1}{2\sigma} y f'(t) \partial_x+ f(t) \partial_y  + \tfrac{1}{2\sigma} f''(t) \partial_u, 
\label{u.gen.y-transl}
\end{align}
while $\X^u_4=\X^u_5 = 0$. 
These generators comprise the local infinitesimal point symmetries 
admitted by the K-KP equation \eqref{KKPeqn.scal}. 

The point transformation groups determined by 
the infinitesimal symmetries \eqref{t-transl}--\eqref{shift2} 
can be obtained in an explicit form by 
integration of the corresponding vector fields,
which is equivalent to the action of $\exp(\epsilon\X)$ on $(t,x,y,v)$. 
Symmetries \eqref{t-transl}, \eqref{shift1} and \eqref{shift2}
directly yield, for the components that are not the identity, 
\begin{equation}
\X_1:\quad
t\to t +\epsilon
\end{equation}
which is a time-translation,
and 
\begin{equation}
\begin{aligned}
& \X_4:\quad
v\to v+\epsilon f(t)
\\
& \X_5:\quad
v\to v+\epsilon y f(t)
\end{aligned}
\end{equation}
which are time-dependent shifts. 
For the other two symmetries \eqref{gen.x-transl} and \eqref{gen.y-transl}, 
the transformation on $v$ is complicated, 
whereas the prolonged transformation on $u$ is much simpler: 
\begin{equation}\label{gen.x-trans.group}
\X^u_2:\quad
x\to x +\epsilon f(t), 
\quad
u\to u + \epsilon f'(t)
\end{equation}
and 
\begin{equation}\label{gen.y-trans.group}
\X^u_3:\quad
x\to x + \epsilon \tfrac{1}{2\sigma} y f'(t) + \epsilon^2 \tfrac{1}{4\sigma} f(t) f'(t) ,
\quad
y\to y+\epsilon f(t),
\quad
u\to u + \epsilon \tfrac{1}{2\sigma} f''(t)
\end{equation}
which are, respectively a generalized $x$-translation and a generalized $y$-translation. 
We consider a few special cases. 

Recall that in the physical context of shallow water waves, 
$u$ has the interpretation of the wave speed. 
Firstly, 
for $f=1$, 
we see that the transformations \eqref{gen.x-trans.group} and \eqref{gen.y-trans.group}
reduce to ordinary translations on $x$ and $y$. 
Secondly, for $f=t$, 
the transformation \eqref{gen.x-trans.group} becomes
a Galilean boost in the $x$ direction with speed $\epsilon$; 
in contrast, 
the transformation \eqref{gen.y-trans.group} with arbitrary fixed $t$
describes a moving reference frame whose velocity has components 
$(\tfrac{1}{4\sigma} \epsilon^2,\epsilon)$ in fixed coordinates $(x,y)$,
and whose position is also shifted by 
$\tfrac{1}{2\sigma}y \epsilon$ in the $x$ direction. 
Lastly, for $f=t^2$, 
the transformation \eqref{gen.x-trans.group} is a boost 
with constant acceleration $\epsilon$ in the $x$ direction,
while the transformation \eqref{gen.y-trans.group} 
describes a moving reference frame with velocity 
$(\tfrac{1}{\sigma} y\epsilon + \tfrac{3}{2\sigma}\epsilon^2 t^2,2\epsilon t)$
and non-constant acceleration $(\tfrac{3}{\sigma}\epsilon^2 t,2\epsilon)$.

\section{Conservation laws}

Conservation laws are of basic importance because
they provide physical, conserved quantities for all solutions $u(x,t)$,
as well as conserved norms which are useful in analysis of solutions, 
and they can be used to check the accuracy of numerical solution methods.
A general discussion of conservation laws and their applications to PDEs 
can be found in \Ref{Olv-book,BCA-book,Anc-review}.

A local conservation law of the potential K-KP equation \eqref{KKPpoteqn}
is a continuity equation
\begin{equation}\label{conslaw}
D_t T + D_x \Phi^x + D_y \Phi^y  =0
\end{equation}
holding for all solutions of equation \eqref{KKPpoteqn}, 
where $T$ is the conserved density and $(\Phi^x,\Phi^y)$ is the spatial flux. 
These three functions depend on $t$, $x$, $y$, $v$, and derivatives of $v$, 
with $v_{tx}$ and its derivatives being eliminated through equation \eqref{KKPpoteqn}.

If $T=D_x\Theta^x + D_y\Theta^y$ and $(\Phi^x,\Phi^y)=-D_t(\Theta^x,\Theta^y)$ 
hold for all solutions $v(x,y,t)$,
where the functions $(\Theta^x,\Theta^y)$ 
depend on $t$, $x$, $y$ $v$, and derivatives of $v$,
then the continuity equation \eqref{conslaw} becomes an identity.
Conservation laws of this form are called locally trivial 
because they contain no useful about solutions $v(x,y,t)$. 
Accordingly, two conservation laws are considered to be locally equivalent
if they differ by a locally trivial conservation law.

The global form of a non-trivial conservation law is given by the balance equation
\begin{equation}\label{globalconslaw}
\frac{d}{dt}\int_{\Omega} T\, dA = -\oint_{\partial\Omega} (\Phi^x,\Phi^y)\cdot \hat n\, ds
\end{equation}
where $\Omega\subseteq\Rnum^2$ is any fixed spatial domain,
and $\partial\Omega$ is its boundary curve with unit outward normal $\hat n$. 
In $(x,y)$ coordinates, 
$dA = dx\,dy$ is the area element and $\hat n\,ds = (dy,-dx)$ is the flux element, 
with $ds$ being the line element. 
A balance equation \eqref{globalconslaw} states that the rate of change of 
the density integral on $\Omega$ 
\begin{equation}\label{cons.integral}
C = \int_{\Omega} T\, dA 
\end{equation}
is equal to the negative of the net outward flux passing through $\partial\Omega$ 
as measured by the flux integral. 
The conserved integral \eqref{cons.integral} is time-independent 
for all solutions $v(x,y,t)$ when the net flux vanishes. 
In the case of $\Omega=\Rnum^2$, the boundary $\partial\Omega$ is taken to be the limit of a circle as its radius goes to infinity. 
Time-independence will then hold for solutions that have sufficient asymptotic spatial decay. 

Every local conservation law can be expressed in an equivalent, characteristic form
(analogous to the evolutionary form for symmetries) 
which is given by a divergence identity
\begin{equation}\label{char-eqn}
D_t \tilde T +D_x \tilde \Phi^x +D_y \tilde \Phi^y = 
( v_{tx} + v_x v_{xx} + \beta v_{xxxx} + v_{xxxxxx} - \sigma v_{yy} ) Q
\end{equation}
holding off of solutions of the potential K-KP equation \eqref{KKPpoteqn},
where 
$\tilde T= T +D_x\Theta^x +D_y\Theta^y$ 
and
$(\tilde \Phi^x,\tilde \Phi^y)= (\Phi^x,\Phi^y) -D_t(\Theta^x,\Theta^y)$
are respectively a conserved density and a spatial flux that are 
locally equivalent to $T$ and $(\Phi^x,\Phi^y)$, 
and where the function $Q$, called a multiplier, 
depends on $t$, $x$, $y$, $v$, and derivatives of $v$, 
such that it is non-singular on solutions $v(x,y,t)$. 

Multipliers are useful because there is a one-to-one correspondence between 
conservation laws (up to local equivalence) 
and multipliers evaluated on solutions of the potential K-KP equation \eqref{KKPpoteqn}.
This correspondence is readily established when $T$, $\Phi^x$, $\Phi^y$ are taken 
without loss of generality to have no dependence on a leading derivative 
$v_{xxxxx}$ or $v_{yy}$ of equation \eqref{KKPpoteqn} 
and all differential consequences of the leading derivative,
as shown by general results known for PDEs with a generalized Cauchy-Kovalevskaya form \cite{Olv-book,BCA-book,Anc-review}.

The Hamiltonian structure \eqref{hamileqn} indicates that the potential K-KP equation
possesses a corresponding Lagrangian formulation
\begin{equation}\label{ELpoteqn}
E_v(L) = v_{tx} + v_x v_{xx} + \beta v_{xxxx} + v_{xxxxxx} -\sigma v_{yy} =0
\end{equation}
with $E_v$ denoting the Euler operator \cite{Olv-book}. 
The Lagrangian is straightforwardly found to be given by 
\begin{equation}\label{lagr}
L = -\tfrac{1}{2} v_t v_x -\tfrac{1}{6} v_x{}^3 + \tfrac{1}{2} \beta v_{xx}{}^2 - \tfrac{1}{2} v_{xxx}{}^2 + \tfrac{1}{2} \sigma v_y{}^2 
\end{equation}
modulo an arbitrary total space-time divergence. 
Noether's theorem states that there is a one-to-one correspondence
between multipliers and the characteristic functions of infinitesimal variational symmetries. 

Recall that an infinitesimal variational symmetry is a generator 
\begin{equation}\label{P.generator}
\hat\X = P\partial_v
\end{equation}
whose prolongation leaves the Lagrangian invariant up to a total space-time divergence, 
\begin{equation}\label{Aterms}
\pr\hat\X(L) = D_t A^t + D_x A^x + D_y A^y
\end{equation}
where $P$ and $A^t$, $A^x$, $A^y$ are functions of 
$t$, $x$, $y$, $v$, and derivatives of $v$. 
The variation $\pr\hat\X(L)$ also satisfies the Euler-Lagrange identity 
\begin{equation}\label{Bterms}
\pr\X(L) = E_v(L) P + D_t B^t + D_x B^x + D_y B^y
\end{equation}
where $B^t$, $B^x$, $B^y$ are functions of $t$, $x$, $y$, $v$, and derivatives of $v$
(which can be found from an explicit formula when needed). 
Equating the variations \eqref{Aterms} and \eqref{Bterms}
shows that $P$ is a multiplier 
\begin{equation}\label{noether.multr}
E_v(L) P = D_t T + D_x \Phi^x + D_y \Phi^y
\end{equation}
yielding a conservation law \eqref{conslaw}, 
with the conserved density and the spatial flux
being given by 
\begin{equation}\label{noether.conslaw}
T=A^t -B^t,
\quad
(\Phi^x,\Phi^y)=(A^x -B^x,A^y -B^y).
\end{equation}

Thus, the Noether correspondence for the potential K-KP equation \eqref{KKPpoteqn} is
\begin{equation}\label{noether}
P = Q
\end{equation}
where $P$ is the characteristic function in a variational symmetry 
and $Q$ is a multiplier. 
Furthermore, as is well known, variational symmetries are symmetries,
because invariance of the Lagrangian implies that the extremals of the Lagrangian 
are also preserved, and therefore
$\hat\X$ will also leave invariant the Euler-Lagrange equation \eqref{ELpoteqn}. 
Then the multiplier equation \eqref{noether.multr}
shows that the condition for a symmetry to be a variational symmetry is given by 
\begin{equation}\label{multr.cond}
E_v( E_v(L) P ) = 0 
\end{equation}
which follows from the property that the Euler operator annihilates a function 
iff it is a total space-time divergence. 

For a given variational symmetry, 
the resulting conserved density and spatial flux (up to local equivalence) 
can be obtained by several methods: 
use of the formula \eqref{noether.conslaw} if expressions 
for $A^t$, $A^x$, $A^y$ and $B^t$, $B^x$, $B^y$ are known 
from equations \eqref{Aterms} and \eqref{Bterms}; 
integration by parts on the multiplier equation \eqref{noether.multr};
a homotopy integral formula applied to the Euler-Lagrange identity \eqref{Bterms}. 
See \Ref{Anc-review} for more details of the methods. 

To apply the variational symmetry condition \eqref{multr.cond} 
to the infinitesimal Lie point symmetries \eqref{t-transl}-\eqref{shift2},
we first obtain their characteristic functions: 
\begin{align}
& 
P_1=-v_t,
\label{P.t-transl}\\
&
P_2 =-v_xf(t)  - ( x f'(t) + \tfrac{1}{2\sigma} y^2 f''(t) ), 
\label{P.gen.x-transl}\\
&
P_3 = -\tfrac{1}{2\sigma} y v_x f'(t)  -v_y f(t)  -  ( \tfrac{1}{2\sigma} y (x f''(t) + \tfrac{1}{6\sigma}y^2 f'''(t)) ),
\label{P.gen.y-transl}\\
&
P_4=f(t), 
\label{P. shift1}\\
&
P_5 = y f(t).
\label{P.shift2}
\end{align}
Next, after substituting these expressions into the condition \eqref{multr.cond}, 
we find that it holds for each one,
and hence all of them are variational symmetries. 
Then we use the integration by parts method to obtain the resulting conservation laws.

\begin{theorem}
The conservation laws of the K-KP potential equation \eqref{KKPpoteqn}
arising from Noether's theorem applied to the variational Lie point symmetries \eqref{t-transl}--\eqref{shift2}
are given by:
\begin{subequations}\label{P1.conslaw}
\begin{flalign}
& 
T_1=\tfrac{1}{2}(v_{xxx}^2-\beta v_{xx}^2+\tfrac{1}{3}v_x^3-\sigma v_y^2),
&\\
&
\Phi^x_1 = (\beta v_{xx}+v_{xxxx})v_{tx}-\tfrac{1}{2}v_x^2-(\beta v_{xxx}+v_{xxxxx}+\tfrac{1}{2}v_x^2)v_t-v_{txx}v_{xxx}, 
&\\
&
\Phi^y_1 = \sigma v_t v_y;
&
\end{flalign}
\end{subequations}
\begin{subequations}\label{P2.conslaw}
\begin{flalign}
& 
T_2=\tfrac{1}{2} v_x^2 f(t) +  v f'(t) , 
&\\
&
\begin{aligned}
\Phi^x_2 & = (\tfrac{1}{2} \sigma v_y^2 + \beta v_x v_{xxx}-\tfrac{1}{2} \beta v_{xx}^2 + \tfrac{1}{3} v_x^3 + v_{xxxxx}v_x - v_{xx}v_{xxxx}+\tfrac{1}{2}v_{xxx}^2) f(t)
\\&\qquad
- ( x (\beta v_{xxx} + v_{xxxxx} + \tfrac{1}{2}v_x^2 + v_t) -(\beta v_{xx}+ v_{xxxx}) ) f'(t)
\\&\qquad
 - \tfrac{1}{2\sigma} y^2  (\beta v_{xxx}  + v_{xxxxx} + v_x^2 + v_t)  f''(t) ,
\end{aligned}
&\\
&
\Phi^y_2 = -\sigma f(t)  v_x v_y + \sigma x f'(t) v_y + (\tfrac{1}{2} y^2 v_y - y v) f''(t);
&
\end{flalign}
\end{subequations}
\begin{subequations}\label{P3.conslaw}
\begin{flalign}
& 
T_3=\tfrac{1}{2 }(v_x v_y f(t) + \tfrac{1}{2\sigma} y v_x^2 f'(t) + \tfrac{1}{\sigma} y v f''(t) ),
&\\
&
\begin{aligned}
\Phi^x_3 & = (\tfrac{1}{2} v_t v_y + (\beta v_{xxx}+\tfrac{1}{2} v_x^2 + v_{xxxxx}) v_y - (\beta v_{xx} + v_{xxxx} ) v_{xy} + v_{xxx} v_{xxy}) f(t)
\\&\qquad
+ ( y ( \tfrac{1}{4} v_y^2 + \tfrac{1}{2\sigma} (\beta v_x v_{xxx} - \tfrac{1}{2} \beta v_{xx}^2 +\tfrac{1}{3} v_x^3 + v_{xxxxx} v_x - v_{xx}v_{xxxx}+\tfrac{1}{2}v_{xxx}^2) ) f'(t)
\\&\qquad
- \tfrac{1}{2\sigma} (  x y (\beta v_{xxx} + \tfrac{1}{2} v_x^2 + v_{xxxxx} + v_t) - y ( \beta v_{xx}+v_{xxxx} )  ) f''(t)
\\&\qquad
- \tfrac{1}{12\sigma^2} y^3 (\beta v_{xxx} + \tfrac{1}{2}v_x^2 + v_{xxxxx} + v_t ) f'''(t) ,
\end{aligned}
&\\
&
\begin{aligned}
\Phi^y_3 & = -\tfrac{1}{2} ((v_x v_t + \sigma v_y^2  - \beta v_{xx}^2 + \tfrac{1}{3} v_x^3 + v_{xxx}^2) f(t)  + y  f'(t) v_x v_y - x (y v_y - v) f''(t) 
\\&\qquad
- \tfrac{1}{2\sigma}(\tfrac{1}{3} y^3 v_y - y^2 v) f'''(t));
\end{aligned}
&
\end{flalign}
\end{subequations}
\begin{subequations}\label{P4.conslaw}
\begin{flalign}
& 
T_4=0,
&\\
&
\Phi^x_4 = (\beta v_{xxx}+v_{xxxxx}+\tfrac{1}{2}v_x^2+v_t) f(t), 
&\\
&
\Phi^y_4 = - \sigma  v_y f(t) ;
&
\end{flalign}
\end{subequations}
\begin{subequations}\label{P5.conslaw}
\begin{flalign}
& 
T_5=0,
&\\
&
\Phi^x_5 =  y (\beta v_{xxx}+v_{xxxxx}+\tfrac{1}{2}v_x^2+v_t) f(t), 
&\\
&
\Phi^y_5 = \sigma (v - y v_y) f(t),
&
\end{flalign}
\end{subequations}
where $f(t)$ is an arbitrary function. 
\end{theorem}

\subsection{Conserved integrals}

All of the preceding conservation laws hold for the K-KP equation 
in physical form \eqref{KKPeqn.scal} through the relation \eqref{pot} 
between the variables $u$ and $v$. 
The first three \eqref{P1.conslaw}--\eqref{P3.conslaw} yield
conserved integrals
\begin{align}
& 
E = \int_{\Omega} \tfrac{1}{2}( u_{xx}^2 -\beta u_{x}^2 -\sigma (\intx u_y)^2 +\tfrac{1}{3}u^3 )\, dx\,dy,
\label{ener}\\
& 
P^x[f] = \int_{\Omega} ( \tfrac{1}{2} f(t) u^2 -f'(t) x u )\, dx\,dy,
\label{gen.x-mom}\\
&
P^y[f] = \int_{\Omega} \tfrac{1}{2} ( f(t) u \intx u_y + \tfrac{1}{\sigma}y(\tfrac{1}{2} f'(t) u^2 - f''(t) x u) )\, dx\,dy,
\label{gen.y-mom}
\end{align}
where integration by parts has been used in the latter two. 
These quantities have the following main features. 
Firstly, 
they coincide with the conserved integrals known for the KP equation \cite{AncGanRec2018}, 
as the effect of the fifth-order dispersive term in the K-KP equation \eqref{KKPeqn.scal}
arises only in the spatial fluxes. 
Secondly, 
they respectively describe 
energy, generalized $x$-momentum, and generalized $y$-momentum,
in accordance with the nature of the variational symmetries
discussed in the previous section. 
We will discuss more of the meaning of the generalized momenta shortly. 

The final two conservation laws \eqref{P1.conslaw}--\eqref{P3.conslaw} 
yield spatial flux integrals
\begin{align}
& \oint_{\partial\Omega} (\beta u_{xx}+u_{xxxx}+\tfrac{1}{2}u^2+\intx u_t) \, dy + \sigma  \intx u_y \,dx 
=0,
\\
& \oint_{\partial\Omega} y (\beta u_{xx}+u_{xxxx}+\tfrac{1}{2}u^2+\intx u_t)\,dy +\sigma (y \intx u_y -\intx u)\,dx
=0.
\end{align}
These quantities describe vanishing topological charges \cite{AncRec2021}
which are unchanged under continuous deformations of the boundary curve $\partial\Omega$. 
They are higher-order versions of the similar topological charges known for the KP equation. 
If the $\intx u_t$ terms are expressed as a time derivative of line integrals
$\oint_{\partial\Omega} \intx u \, dy$ and $\oint_{\partial\Omega} y \intx u \,dy$, 
and then the divergence theorem is used to write them as 2D volume integrals, 
we obtain conserved integrals
\begin{equation}\label{mass}
M = \int_{\Omega} u\, dx \,dy,
\qquad
M^y = \int_{\Omega} yu\, dx \,dy.
\end{equation}
The first one describes net mass,
and the second one describes the $y$-moment of the mass distribution,
both of which coincide with the same conserved quantities known for the KP equation \cite{AncGanRec2018}. 
These quantities are connected to the meaning of 
the conserved momenta \eqref{gen.x-mom} and  \eqref{gen.y-mom} as follows. 

Firstly, for $f=1$, 
the momenta quantities \eqref{gen.x-mom} and  \eqref{gen.y-mom}  
reduce to the ordinary $x$- and $y$- momenta 
\begin{equation}\label{momenta}
P^x = \int_{\Omega} \tfrac{1}{2} u^2 \, dx\,dy,
\quad
P^y = \int_{\Omega} \tfrac{1}{2} u \intx u_y \, dx\,dy.
\end{equation}
Secondly, for $f=t$, 
the quantity \eqref{gen.x-mom} describes Galilean $x$-momentum 
whose conservation implies the relation 
$M^{x}(t) = \int_{\Omega} x u \, dx\,dy= t P^x +M^{x}(0)$. 
Since total mass $M$ is conserved, 
we conclude that the $x$-center of mass, given by 
$\chi^M(t) = M^x(t)/M = t P^x/M +\chi^M(0)$, 
moves at a constant speed $P^x/M$. 
Similarly, 
the conserved integral \eqref{gen.y-mom} is a Galilean-like momentum, 
which is associated to the moving reference frame described by 
the underlying symmetry transformation \eqref{gen.y-trans.group} 
in the previous section. 
The conservation of this quantity is equivalent to the relation 
$P^{x,y}(t) = \int_{\Omega} \tfrac{1}{2}y u^2 \, dx\,dy = P^{x,y}(0) -2\sigma t P^y$, 
showing that the $y$-center of $x$-momentum 
$\chi^{P^x}(t) = P^{x,y}(t)/P^x = -2\sigma t P^y/P^x +\chi^{P^x}(0)$, 
moves at a constant speed $-2\sigma P^y/P^x$. 

We now state some final results on variational symmetries and conservation laws. 
In principle, variational symmetries -- and hence multipliers -- could comprise 
more than just Lie point symmetries. 
Since the K-KP equation \eqref{KKPeqn.scal} is of even order,
we expect all multipliers to be of odd order,
in which case the next possible higher multiplier 
after the ones given by variational Lie point symmetries 
would have order three
\begin{equation}\label{low-order.Q}
Q=P(t,x,y,v,\partial v,\partial^2 v,\partial^3 v)
\end{equation}
where $\partial=(\partial_t,\partial_x,\partial_y)$, 
$\partial^2=(\partial_t^2,\partial_x^2,\partial_y^2,\partial_t\partial_x,\partial_t\partial_y,
\partial_x\partial_y)$, 
and so on. 
The corresponding conserved density would be of order four in $v$, 
which would yield a conserved integral of order three in $u$. 

The determining equation \eqref{noether.multr} for such multipliers \eqref{low-order.Q}
can be split with respect to derivatives of $v$, 
yielding an overdetermined system of equations to find $Q$ directly. 
It is straightforward using Maple to set up and solve this system. 
We find that the only solutions are the multipliers given by 
the characteristics \eqref{P.t-transl}--\eqref{P.shift2} of the variational symmetries. 
This means that the K-KP equation does not possess 
any multipliers of order two or three, 
and likewise no variational symmetries of order two or three exist. 
Consequently, the K-KP equation does not possess conserved integrals 
whose density is of order four in $v$, or order three in $u$.

\section{Variational structure}

For any Hamiltonian evolution equation,
there is a well-known \cite{Olv-book} analog of Noether's theorem
which produces a symmetry from each admitted conserved integral. 
It can be formulated for the K-KP equation \eqref{KKPeqn.scal}
by the explicit relation
\begin{equation}\label{hamil.noether}
P^u = \Hop(\delta C/\delta u) = D_x E_u(T)
\end{equation}
involving the characteristic function $P^u$ of an infinitesimal symmetry generator 
\begin{equation}\label{P.generator.u}
\hat\X^u = P^u\partial_u
\end{equation}
and the density $T$ of a conserved integral \eqref{cons.integral}. 
Furthermore, because equation \eqref{KKPeqn.scal} is an evolution equation, 
the expression 
\begin{equation}\label{QTrel.u}
Q^u =E_u(T)
\end{equation}
is a conservation law multiplier, 
\begin{equation}
Q^u G = D_t T + D_x \Phi^x + D_y \Phi^y
\end{equation}
where 
$G =u_t + \alpha u u_x +\beta u_{xxx}+\gamma u_{xxxxx} -\sigma \partial^{-1} u_{yy}$. 
Hence, relation \eqref{hamil.noether} can be expressed as 
\begin{equation}\label{PQ.rel}
P^u = D_x Q^u.
\end{equation}
This correspondence is one way: every conserved integral yields an infinitesimal symmetry.
The converse holds iff the symmetry has the Hamiltonian form \eqref{hamil.noether}. 

The same correspondence can be derived from Noether's theorem
applied to the Lagrangian \eqref{lagr} 
for the K-KP equation in potential form \eqref{KKPpoteqn}. 
Variational symmetries \eqref{P.generator} of this equation 
have a one-to-one correspondence \eqref{noether} with multipliers. 
When the conservation law produced by a variational symmetry 
has $T$ and $(\Phi^x,\Phi^y)$ given by functions of only $v_x=u$ and its derivatives, 
then the multiplier will have the form \eqref{QTrel.u}, 
where 
\begin{equation}\label{Qrel.u}
Q^u=Q
\end{equation}
holds due to the equality $E_v(L)=G\big|_{u=v_x}$. 
Now we observe that a variational symmetry \eqref{P.generator} will project to 
a symmetry \eqref{P.generator.u} of the K-KP equation \eqref{KKPeqn.scal} 
through the relation 
\begin{equation}\label{Prel.u}
P^u=D_x P.
\end{equation}
Then the Noether correspondence \eqref{noether} 
combined with the preceding two relations \eqref{Qrel.u} and \eqref{Prel.u}
yields the Hamiltonian correspondence \eqref{PQ.rel}. 

All three Lie point symmetries \eqref{u.t-transl}--\eqref{u.gen.y-transl} 
are Hamiltonian symmetries, since they arise from variational point symmetries of
the potential K-KP equation \eqref{KKPpoteqn}.

\section{Line soliton solutions}

A line soliton is a solitary travelling wave of the form 
\begin{equation}\label{linesoliton}
u = U(\xi),
\quad
\xi = x+\mu y - \nu t
\end{equation}
in terms of parameters $\mu$ and $\nu$ which determine 
the speed and the direction of the wave (see e.g. \cite{AncGanRec2018,AncGanRec2021}). 
This type of solution is group-invariant with respect to the translation symmetries
$\X^u = \partial_t +\nu\partial_x$ and $\X^u = \partial_y -\mu\partial_x$,
which are linear combinations of the generators \eqref{u.t-transl}, 
and both \eqref{u.gen.x-transl} and \eqref{u.gen.y-transl} with $f=1$. 

Substitution of expression \eqref{linesoliton} into the K-KP equation \eqref{KKPeqn.scal} 
yields a fifth-order nonlinear ODE
\begin{equation}\label{tw.ODE}
UU' - \kappa U' + \beta U''' + U''''' = 0
\end{equation}
where
\begin{equation}
\kappa =\sigma\mu^2 +\nu.
\end{equation}
We will be interested in solutions such that $U$ asymptotically approaches a constant 
for large $|\xi|$,
namely, solitary line waves on a constant background. 
This will include the case when the background is zero. 

Integration of the ODE \eqref{tw.ODE} gives 
\begin{equation}\label{ODE.4thorder}
\tfrac{1}{2} U^2 - \kappa U  +\beta U'' + U'''' = C_1 
\end{equation}
and use of the integrating factor $U'$ gives 
\begin{equation}\label{ODE.3rdorder}
\tfrac{1}{6} U^3 - \tfrac{1}{2}\kappa U^2  +\tfrac{1}{2}\beta U'{}^2 
+ U'U''' -\tfrac{1}{2} U''{}^2 = C_1 U + C_2
\end{equation}
where $C_1$, $C_2$ are constants. 
No other simple integrating factors appear to exist. 
It is worth noting that the fourth-order ODE \eqref{ODE.4thorder} 
is an Euler-Lagrange equation $\delta L^U/\delta U =0$
for the Lagrangian 
\begin{equation}
L^U = \tfrac{1}{2}( U''{}^2  -\beta U'{}^2 - \kappa U^2 +\tfrac{1}{3} U^3 ) - C_1 U .
\end{equation}
The terms in this Lagrangian can be expressed as
$L^U = E^U - \nu P^U -C_1 M$ 
where 
\begin{equation}
E^U = \tfrac{1}{2}( U''{}^2 -\beta U'{}^2 -\sigma \mu^2 U^2 +\tfrac{1}{3}U^3 ),
\quad
P^U = \tfrac{1}{2} U^2,
\quad
M^U = U
\end{equation}
are respectively the densities in the conserved integrals for 
energy \eqref{ener}, $x$-momentum \eqref{momenta}, and mass \eqref{mass}, 
evaluated for line waves \eqref{linesoliton}. 

We will now seek explicit sech-type solutions for $U(\xi)$
by applying the $\tanh$ method \cite{Hereman}. 
Consider a polynomial 
\begin{equation}\label{tanh.series}
U(\xi)= a_0 + \sum_{j=1}^{n} a_j \tanh(m \xi)^j
\end{equation}
where $m$, $n$, and $a_j$ ($j=0,1,\ldots,n$) are unknown constants. 
The degree $n$ is determined by balancing the respective powers of $\tanh(m\xi)$ 
that occur in the highest nonlinear term, $U^3$, 
and in the highest derivative term, $U'U'''$, in the ODE \eqref{ODE.3rdorder}:
$3n=2n+4$. 
This yields
\begin{equation}\label{tanh.degree}
n=4.
\end{equation}
Since the highest degree in the polynomial \eqref{tanh.series} is even, 
we can restrict the remaining terms to also have an even degree, 
whereby $U(\xi)$ will be an even function of $\xi$ as is typical for a solitary wave. 
Moreover, via the identity $\tanh^2 = 1-\sech^2$, 
$U(\xi)$ then has a sech-form: 
\begin{equation}\label{sech.series}
U(\xi)= \tilde a_0 + \tilde a_1 \sech(m \xi)^2 + \tilde a_2 \sech(m \xi)^4
\end{equation}
with unknown constants 
$\tilde a_0$ $(=a_0 -a_2 -a_4)$, $\tilde a_1$ $(=-a_2 -2 a_4)$ , $\tilde a_2$ $(=a_4)$,  
and $m$. 

We next substitute this ansatz \eqref{sech.series} into the ODE \eqref{ODE.3rdorder}, 
rewrite the resulting expression in terms of $\tanh(m\xi)$, 
and separately put to zero the coefficient of each power of $\tanh(m\xi)$. 
This yields an overdetermined nonlinear system 
which consists of 7 algebraic equations involving 
$m$, $\tilde a_0$, $\tilde a_1$, $\tilde a_2$, $C_1$, $C_2$, and also $\kappa$, $\beta$. 
It is straightforward to set up and solve this system in Maple by use of the `rifsimp' command. 
A single real solution is obtained: 
\begin{equation}\label{sol.constants}
\begin{aligned}
&
(2m)^2 + \tfrac{1}{13}\beta =0,
\\
& 
\tilde a_0 = \kappa +36 (2m)^4, 
\quad
\tilde a_2=0, 
\quad 
\tilde a_4=-105 (2m)^4, 
\\
&
C_1=\tfrac{1}{2}(\kappa -36(2m)^4)(\kappa+ 36(2m)^4),
\quad 
C_2 =\tfrac{1}{6} (\kappa - 72(2m)^4)(\kappa+ 36(2m)^4).
\end{aligned}
\end{equation}
Substitution of this algebraic solution into the ansatz \eqref{sech.series}
gives the following result. 

\begin{proposition}\label{prop:sech.solns}
All solitary line wave solutions of the K-KP equation \eqref{KKPeqn.scal} 
having a $\sech$ form are given by 
\begin{equation}\label{u.line.soliton}
\begin{aligned}
& u = U(\xi) = p - q\sech(\tfrac{1}{2}r\xi)^4,
\\& 
q= 105 r^4,
\quad
p =\sigma\mu^2 +\nu +36 r^4, 
\quad
r= \sqrt{\tfrac{1}{13}|\beta|},
\quad
\beta <0.
\end{aligned}
\end{equation}
These solutions describe dark solitary waves on a background $p$. 
\end{proposition}

Note the condition $\beta<0$ means that third-order and fifth-order dispersion terms 
in the K-KP equation have opposite signs. 

\begin{figure}[h]
\centering
\includegraphics[width=.48\textwidth]{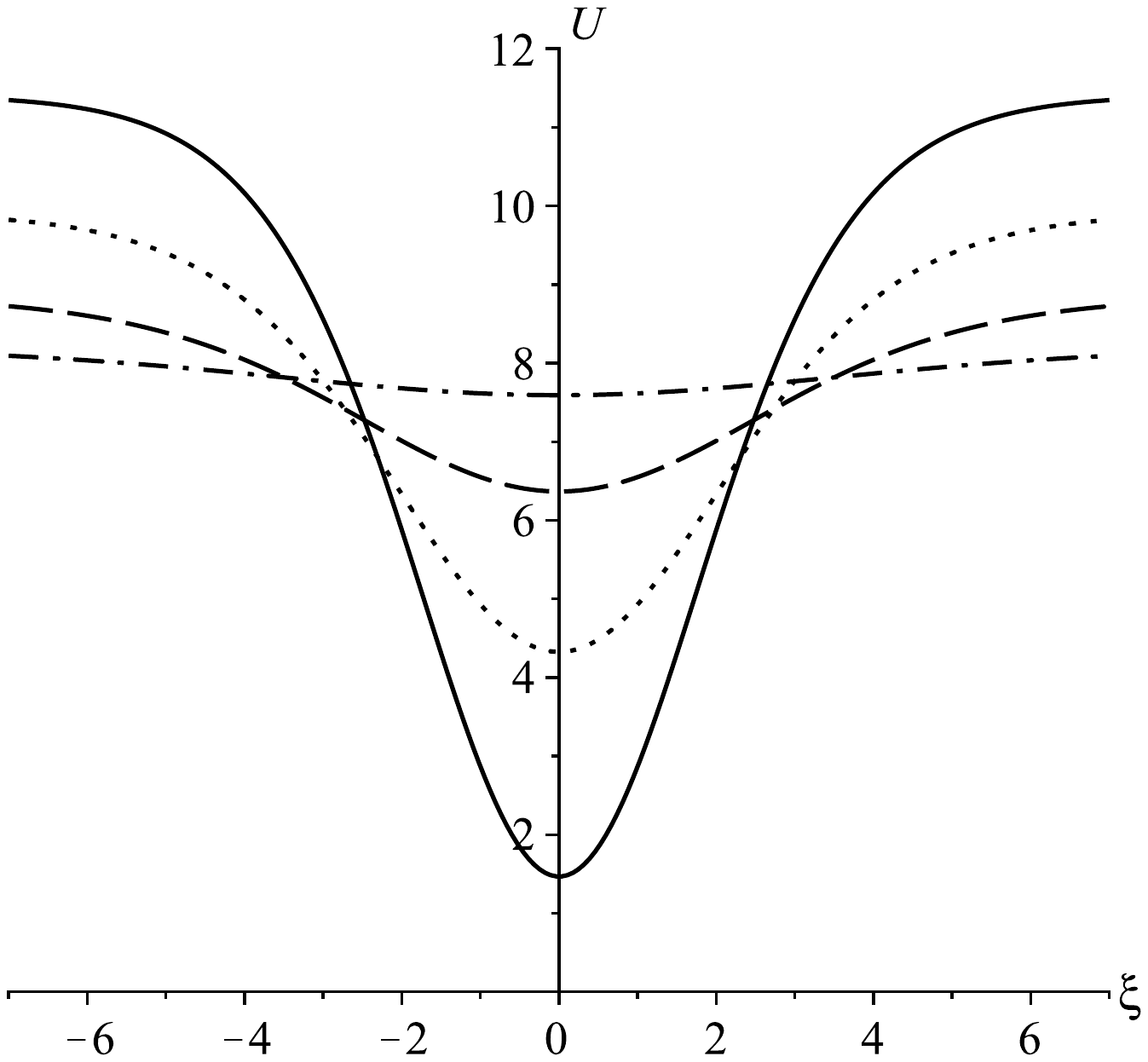}
\quad
\includegraphics[width=.48\textwidth]{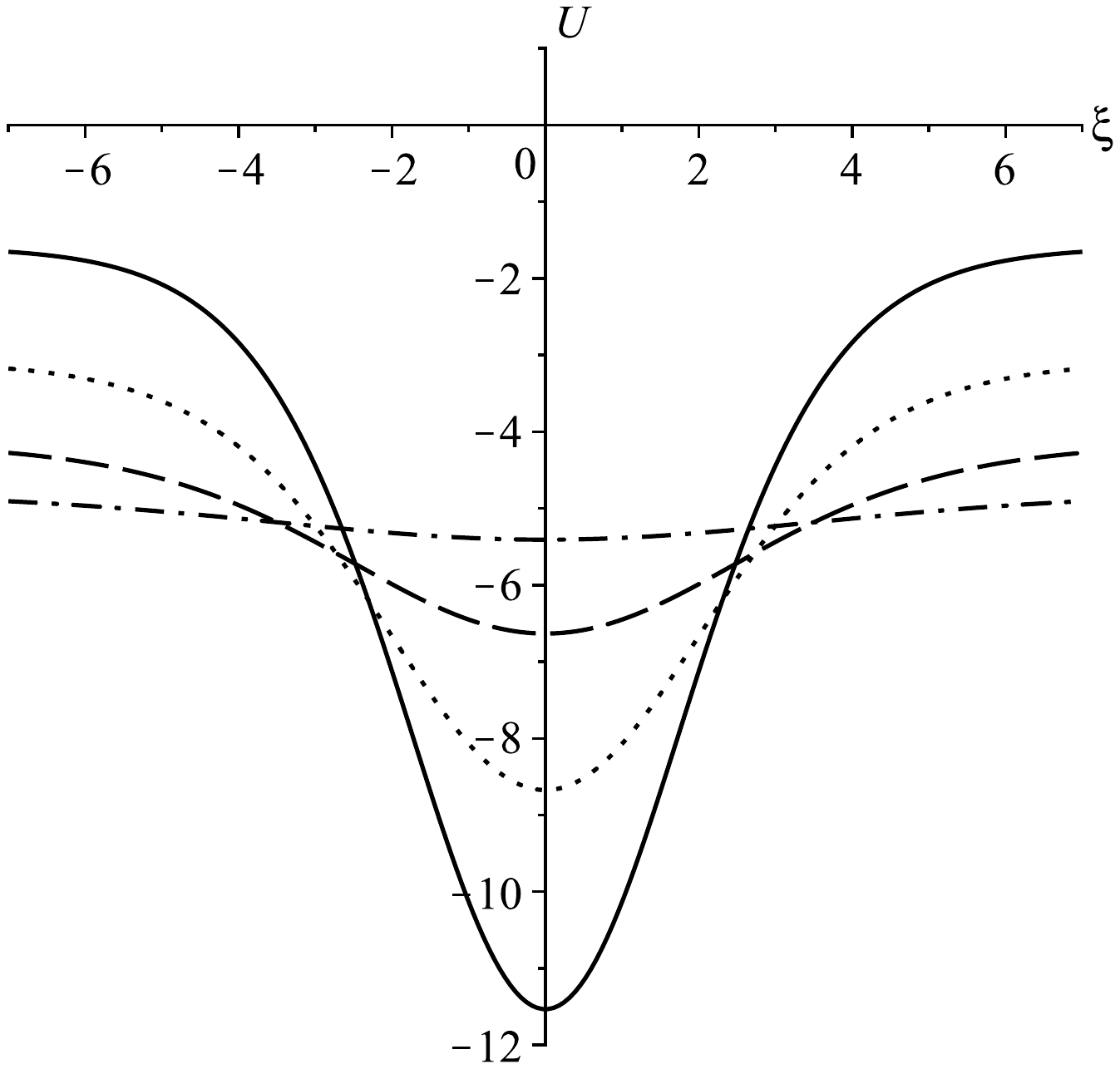}
\caption{Line soliton profile 
for $\beta=-4$ (solid), $-3$ (dot), $-2$ (dash), $-1$ (dot-dash), 
with $\kappa=8$ (left) and $\kappa=-5$ (right).}
\label{fig:background_pos_neg}
\end{figure}

\begin{figure}[h]
\centering
\includegraphics[trim=3cm 5cm 3cm 7cm,clip,width=0.6\textwidth]{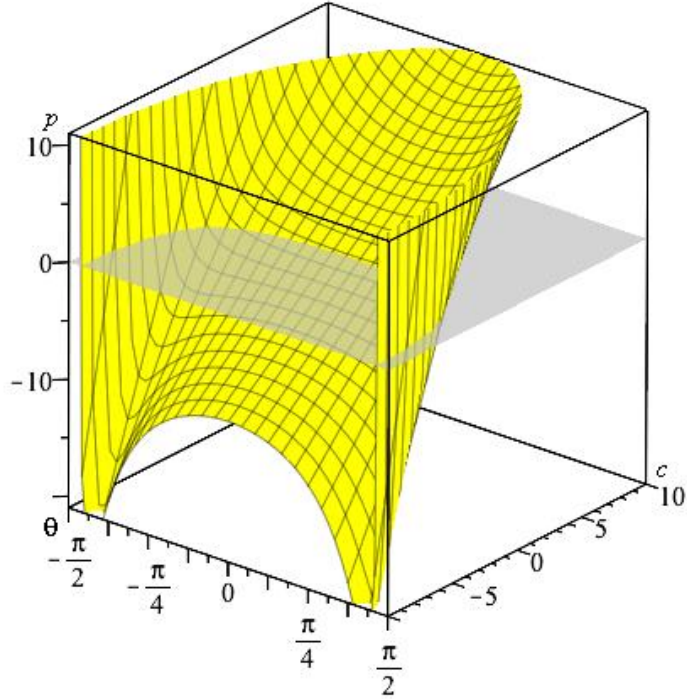}
\caption{Size of background of line soliton profile.}
\label{fig:background_c_theta}
\end{figure}

The line solitons \eqref{u.line.soliton} have the following main properties. 
They have speed $c= \nu/\sqrt{1+\mu^2}$ in the plane $(x,y)$
and have direction angle $\theta=\arctan(\mu)$ with respect to the $x$-axis. 
Their width is proportional to $1/\sqrt{|\beta|}$,
and their depth is proportional to $\beta^2$, 
which are independent of the velocity. 
Hence, narrower waves are taller, but have the same speed. 
See Fig.~\ref{fig:background_pos_neg}. 
The background size $p=\sigma\tan(\theta)^2 + c/|\cos(\theta)| + (\tfrac{6}{13}\beta)^2$
increases with $\beta^2$ with an offset which depends on velocity
and which can be either positive or negative in sign 
depending on $\sigma=\pm 1$ and $c$. 
See Fig.~\ref{fig:background_c_theta}. 

\begin{figure}[h]
\centering
\includegraphics[width=.48\textwidth]{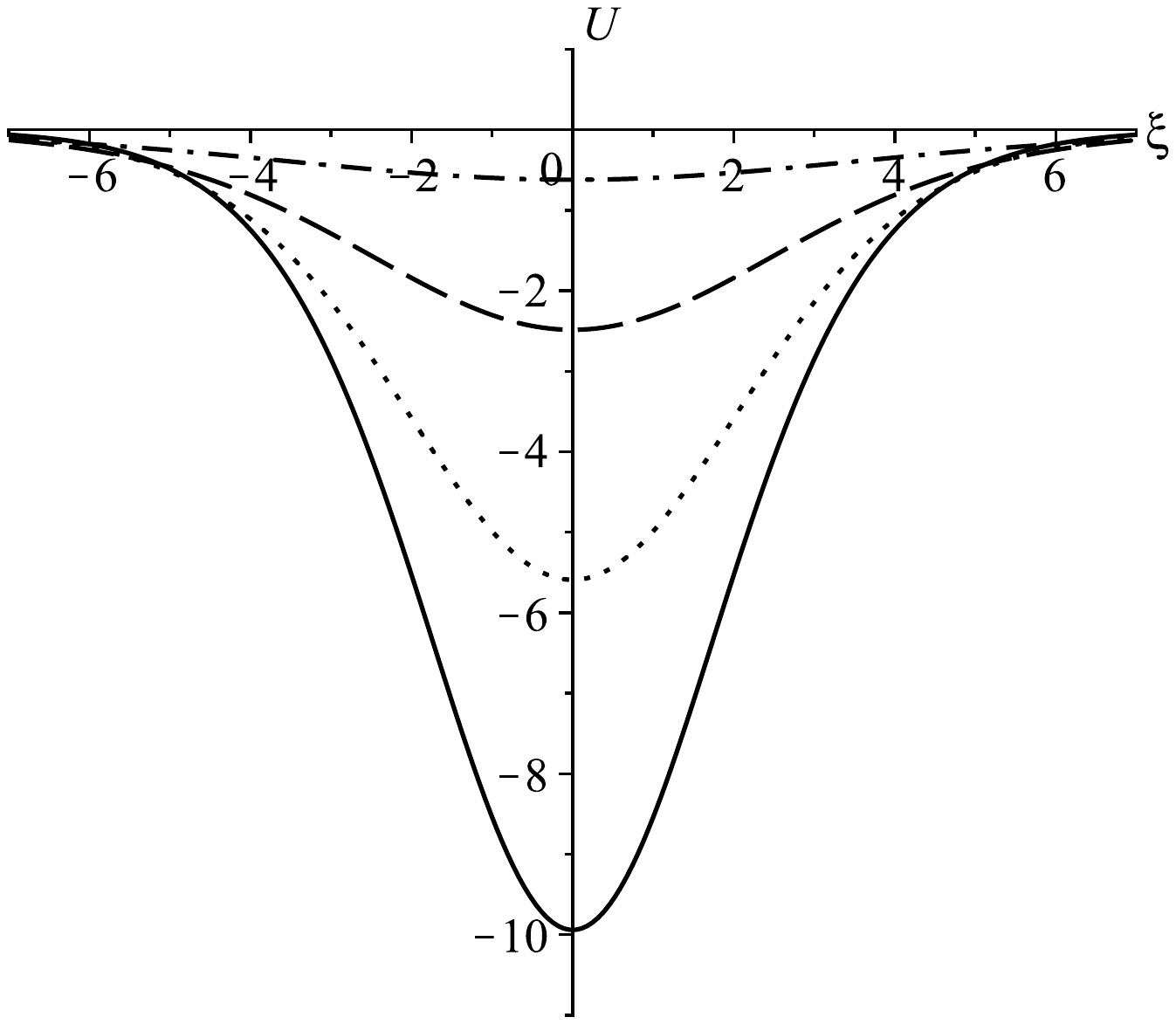}
\caption{Line soliton profile with zero-background 
for $\beta=-4$ (solid), $-3$ (dot), $-2$ (dash), $-1$ (dot-dash).}
\label{fig:background_zero}
\end{figure}

A vanishing background, $p=0$, occurs when the condition 
$\sigma\mu^2 +\nu = - (\tfrac{6}{13}\beta)^2$
holds. 
In terms of terms of speed and direction, this condition is given by 
\begin{equation}\label{kin.cond}
\sigma\tan(\theta)^2 +c/|\cos(\theta)| = -(\tfrac{6}{13}\beta)^2.
\end{equation}
See Fig.~\ref{fig:background_zero}. 

For fixed $\beta<0$, 
we can readily find the region in the kinematic parameter space $(c,\theta)$ 
where equation \eqref{kin.cond} is satisfied. 
Note that $\theta\to\theta \pm \pi$ is equivalent to $c\to -c$. 
Hence, without loss of generality, 
we will allow $c$ to be positive or negative, 
while restricting $-\tfrac{\pi}{2}<\theta<  \tfrac{\pi}{2}$. 

\begin{figure}[h]
\centering
\includegraphics[width=.48\textwidth]{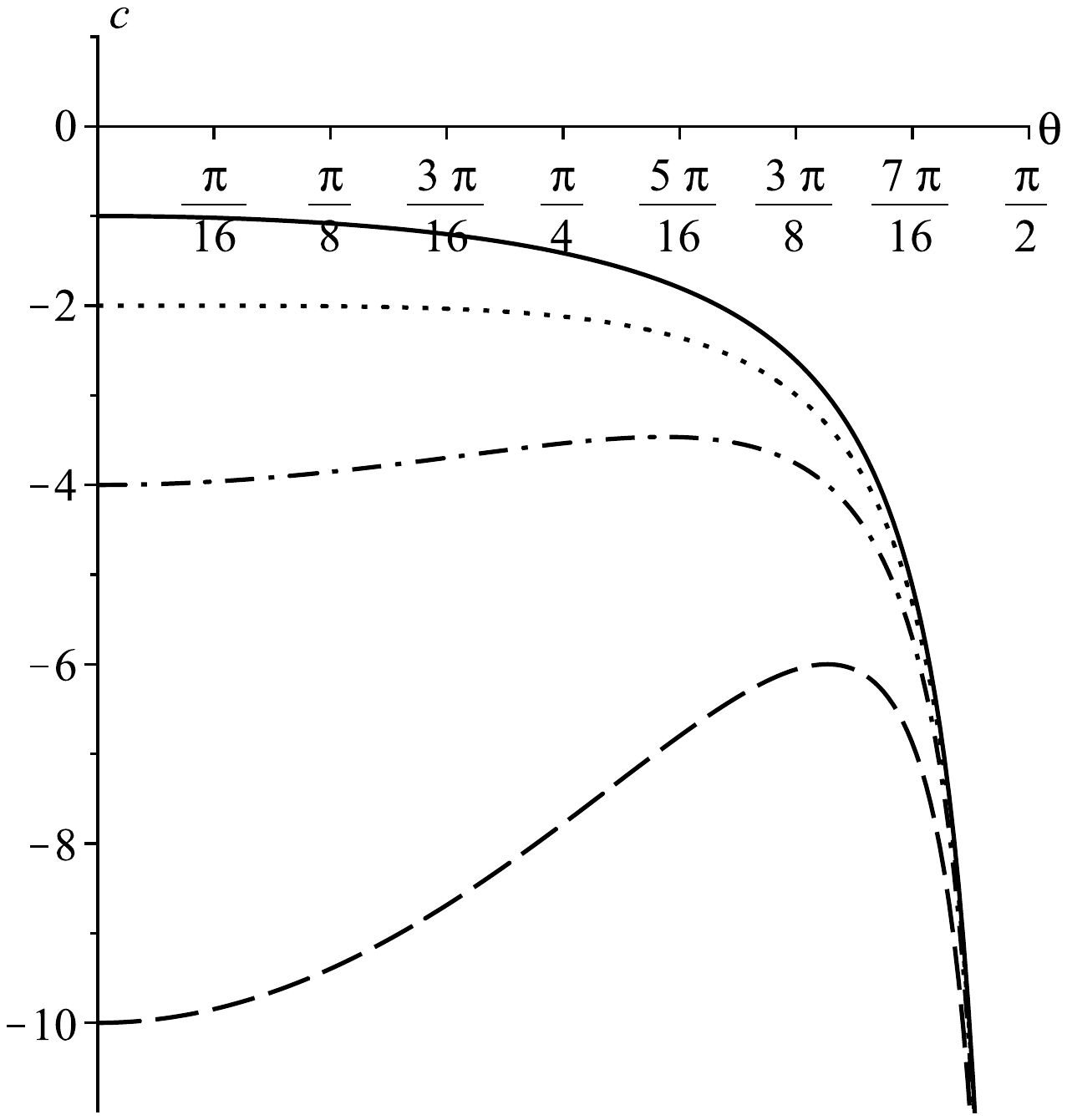}
\quad
\includegraphics[width=.48\textwidth]{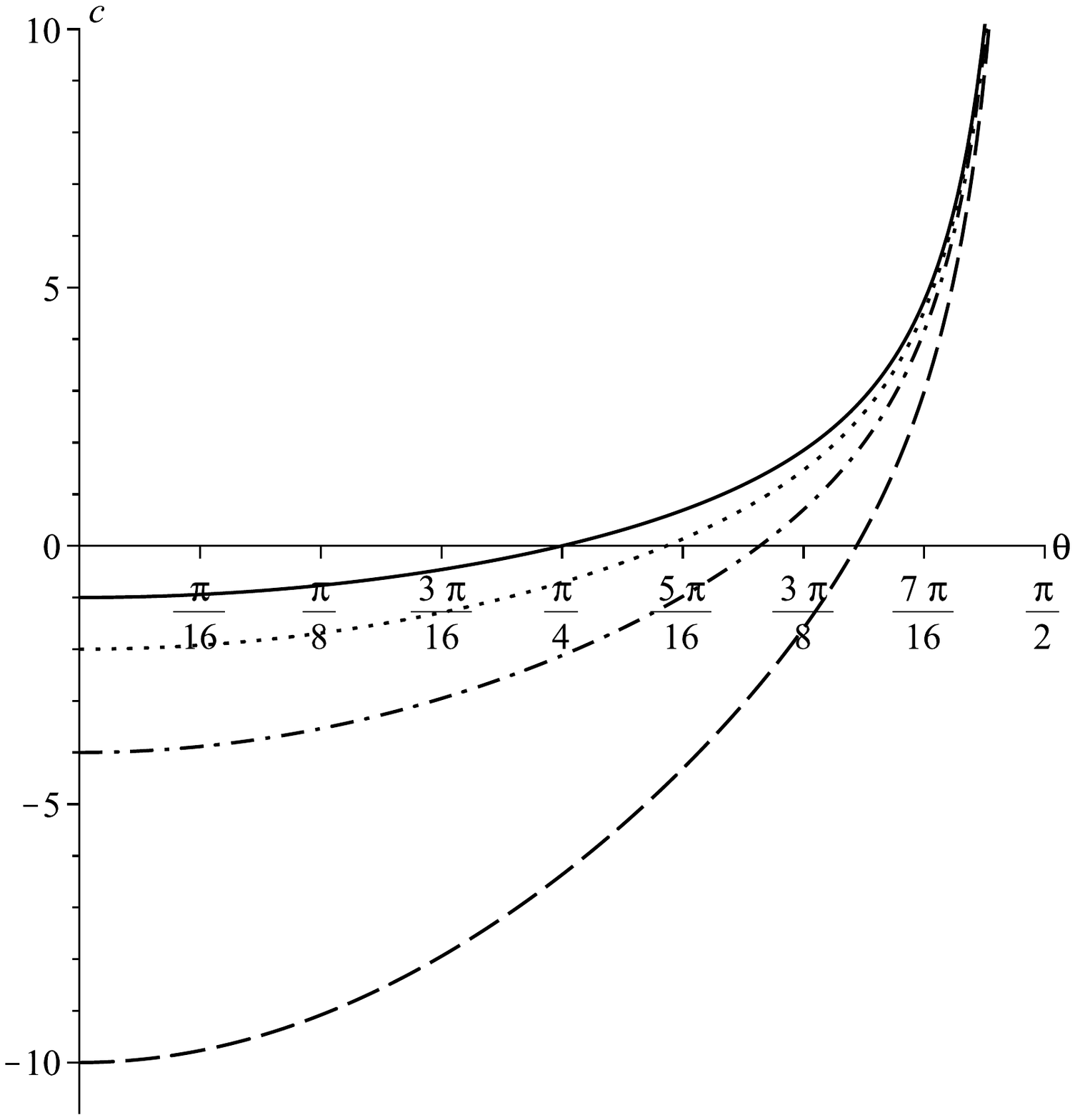}
\caption{Kinematically allowed curve in $(c,\theta)$ for zero-background line solitons 
with $(\tfrac{6}{13}\beta)^2=1$ (solid), $2$ (dot), $4$ (dot-dash), $10$ (dash), 
for $\sigma=1$ (left) and $\sigma=-1$ (right).}
\label{fig:kin.conds}
\end{figure}

First, at a fixed direction angle $\theta$:
the kinematic condition \eqref{kin.cond} yields
\begin{equation}\label{c.cond}
c(\theta) = -(\tfrac{6}{13}\beta)^2 |\cos(\theta)| -\sigma\sin(\theta)^2/|\cos(\theta)|.
\end{equation}
Waves that move in the $x$ direction ($\theta=0$) will be considered later. 
For waves that move transversely ($\theta\neq 0$), 
the shape of the curve \eqref{c.cond} is different in the two cases $\sigma=\pm 1$. 
See Fig.~\ref{fig:kin.conds}. 

In the case $\sigma =1$, 
if $(\tfrac{6}{13}\beta)^2\leq 2$ then 
$c$ decreases without bound as $\theta$ increases to $\pm \pi/2$; 
whereas if $(\tfrac{6}{13}\beta)^2>2$ then 
$c$ increases to a negative maximum 
$c_{\max}=-2\sqrt{(\tfrac{6}{13}\beta)^2-1}$ 
at $|\theta|=\arctan\big(\sqrt{(\tfrac{6}{13}\beta)^2-2}\big)$, 
and subsequently $c$ decreases without bound as $|\theta|$ increases to $\pi/2$. 

In the case $\sigma =-1$, 
$c$ increases with $|\theta|$, 
reaches $0$ at $|\theta|=\arctan\big(\tfrac{6}{13}|\beta|\big)$, 
and continues to increase without bound as $|\theta|$ goes to $\pi/2$. 
Hence, a stationary solitary wave exists 
at the angle $|\theta|=\arctan\big(\tfrac{6}{13}|\beta|\big)$. 

\subsection{Comparison with sech-type solitary waves of the Kawahara equation} 

We now specialize to considering line waves \eqref{linesoliton} 
that move strictly in the $x$-direction:
$\mu =0$, $\nu=c$,
where $c$ is the wave speed. 
The ODE governing these line waves is obtained from 
the fifth-order ODE \eqref{tw.ODE} for transversely-moving line waves 
by putting 
\begin{equation}\label{xdir}
\kappa=c
\end{equation}
which gives
\begin{equation}\label{tw.ODE.xdir}
UU' - c U' + \beta U''' + U''''' = 0 . 
\end{equation}
This ODE has the same form as that for travelling waves of the scaled Kawahara equation 
\begin{equation}\label{Keqn.scal}
u_t + u u_x +\beta u_{xxx} + u_{xxxxx} =0 . 
\end{equation}
In particular, the term in the K-KP equation \eqref{KKPeqn.scal} 
responsible for describing transverse motion --- namely, in the $y$ direction ---
has no effect for line waves moving in the $x$ direction.
Thus, equation \eqref{KKPeqn.scal} is reduced effectively to the Kawahara equation \eqref{Keqn.scal}, 
whereby we immediately see that travelling waves $u=U(\xi)$ with $\xi=x-ct$ 
satisfy the ODE \eqref{tw.ODE.xdir}. 

Integrating this ODE twice then gives the third-order ODE \eqref{ODE.3rdorder} 
with the kinematic condition \eqref{xdir} 
which characterizes waves in the $x$-direction. 
By imposing this condition on the family of solutions given in Proposition~\ref{prop:sech.solns},
we obtain all sech-type solitary wave solutions that move in the $x$-direction: 
\begin{equation}\label{u.soliton.xdir}
\begin{aligned}
& u = U(\xi) = p - q\sech(\tfrac{1}{2}r\xi)^4,
\quad
\xi =x-ct,
\\& 
q= 105 r^4,
\quad
p =c +36 r^4, 
\quad
r= \sqrt{\tfrac{1}{13}|\beta|},
\quad
\beta <0.
\end{aligned}
\end{equation}
These solutions describe dark solitary waves on a background $p$ whose sign 
depends on $c$. 

In the case of zero background, $p=0$, 
note that the solution family explicitly becomes a single solution 
\begin{equation}\label{u.soliton.xdir.zerobackground}
\begin{aligned}
& u = U(\xi) = -\tfrac{105}{169} \beta^2 \sech(\sqrt{\tfrac{1}{52}|\beta|}\xi)^4,
\quad
\xi =x-ct,
\\& 
c = -(\tfrac{6}{13}\beta)^2,
\quad
\beta <0
\end{aligned}
\end{equation}
which is a dark solitary wave having negative speed. 

In the literature on explicit solitary wave solutions of the Kawahara equation,
the zero-background solution \eqref{u.soliton.xdir.zerobackground} 
in a rescaled form is presented in \Ref{Alb}. 
Using the sech-type anstaz \eqref{sech.series} with $\tilde a_0=0$, 
\Ref{Nat} claims to find another sech-type solitary wave with a zero background; 
however, the purported solution can be explicitly verified to not satisfy the governing ODE.
Non-zero background solutions are derived in \Ref{Sir} 
by means of a more general ansatz than \eqref{tanh.series}, 
but only a single sech-type solution is reported; 
it turns out to be a particular case of the family \eqref{u.soliton.xdir}:
$p=\tfrac{72}{169}$, $c=\tfrac{36}{169}$, 
with $\beta=-1$ fixed in the equation.

\section{Concluding remarks}

We have studied several aspects of the K-KP equation \eqref{KKPeqn}:
symmetries, conservation laws and conserved integrals, 
and solitary line wave solutions. 

From a classification of Lie point symmetries,
we have found that these symmetries comprise 
time-translation, plus generalized $x$- and $y$- translations
which depend on an arbitrary function of time. 
When a potential is introduced, 
there are additionally two shifts, which depend on an arbitrary function of time. 
In special cases, the generalized $x$- and $y$- translations reduce to 
ordinary translations, a Galilean boost in the $x$ direction, 
and transformations to moving reference frames 
with either constant velocity or acceleration in the $(x,y)$ plane. 

The use of the potential leads to a Lagrangian formulation,
allowing Noether's theorem to be applied. 
We have shown that all of the Lie point symmetries are variational symmetries
and thus give rise to corresponding conservation laws. 
The conserved quantities defined by these conservation laws describe
energy, plus generalized $x$- and $y$- momenta which depend on an arbitrary function of time. 
These two momenta include, as special cases, 
ordinary $x$-momentum and $y$-momentum, 
a Galilean $x$-momentum, 
and a similar momentum associated with the previous moving reference frames 
in the $(x,y)$ plane. 

We have also derived explicit sech-type line wave solutions 
which describe dark solitary waves on zero and non-zero backgrounds. 
Due to the higher-order dispersion in the K-KP equation, 
these waves exhibit some features which differ from the well known 
line solitons of the KP equation. 
In particular, the size of the background depends on the dispersion ratio 
and on the speed and direction of the waves. 
We have explored the zero-background case and found that 
the wave speed is a function of the direction when the dispersion ratio is fixed. 

These results are new and can be used in the study of physical applications of the K-KP equation. 

An open question that naturally arises is whether the sech-type line wave solutions 
in Proposition~\ref{prop:sech.solns} are stable. 
There are several different notions of stability that have been considered
for line wave solutions of the Kawahara and KP equations: 
perturbative stability with respect to growth in time or in transverse directions; 
orbital stability as a travelling wave; 
global minimizer among all travelling waves.
All of these stability results have been formulated for zero-background solutions.
Stability of solutions having a non-zero background, say $u_0\neq 0$, 
can be approached by applying a shift transformation $\tilde u = u -u_0$ to both 
the solution $u$ and the equation for $u$, 
which then allows the use of standard stability methods 
for zero-background solitary waves. 

In \Ref{Alb}, orbital stability is proven for the zero-background solution \eqref{u.soliton.xdir.zerobackground} 
of the Kawahara equation \eqref{Keqn.scal} in a scaled form. 
This result can be seen in essence to rely only on properties of the ODE for solitary waves, 
and thus it can be adapted in outline to line wave solutions $u=u_0+ \tilde U(\xi)$ 
of the K-KP equation \eqref{KKPeqn.scal} 
with $\tilde U(\xi)$ having asymptotic decay on a background $u_0\neq 0$,
where $\xi$ is the line wave variable \eqref{linesoliton}. 
For simplicity, consider a rescaled form of the K-KP equation \eqref{KKPeqn.scal}, 
with $\beta<0$, corresponding to the notation in \Ref{Alb}:
$u_t + uu_x + \tfrac{13}{420} u_{xxx} - \tfrac{1}{1680} u_{xxxxx} 
+ \epsilon \partial_x^{-1} u_{yy} =0$
where $\epsilon =0$ is the Kawahara case and $\epsilon \neq0$ is the K-KP case. 
When the explicit sech-type solutions \eqref{u.line.soliton} are rescaled 
to satisfy this equation, 
they take the form $\tilde U(\xi) = \sech^4(\xi)$ 
with $u_0= \nu-\epsilon\mu^2 -\tfrac{12}{35}$. 
This family includes the zero-background solution in \Ref{Alb} 
for the case $\epsilon=0$, 
which is given by $u_0=0$, $\nu=c= \tfrac{12}{35}$. 
All of these solutions satisfy the ODE
\begin{equation}\label{tw.ODE.scal}
\tfrac{1}{1680} \tilde U''''  -\tfrac{13}{420}  \tilde U'' = \tfrac{1}{2}\tilde U^2 
\end{equation}
which is equivalent to the ODE \eqref{ODE.4thorder} after a scaling and a shift. 

A key property of this ODE \eqref{tw.ODE.scal} is that 
the linear operator defined by the lefthand side, 
$\mathcal{L} =\tfrac{1}{1680}  (d/d\xi)^4 - \tfrac{13}{420} (d/d\xi)^2$, 
has a positive Fourier transform, 
which is a main condition used in the proof of stability in \Ref{Alb}. 
The other main condition is that, as shown in \Ref{Alb}, 
the value of a certain integral is negative:
$I=\int_{-\infty}^{\infty} \tilde U\psi\, d\xi<0$,
where $\psi(\xi)$ is an $L^2$ function that satisfies $\mathcal{L}\psi = \tilde U$. 
Since this value depends only on the operator and on the expression for $\tilde U$, 
it holds not just for the case $\epsilon=0$ but also for the case $\epsilon\neq0$. 
Therefore, the stability proof for the zero-background solution from \Ref{Alb} 
should extend to the non-zero background solutions. 
This would establish stability of the sech-type line waves \eqref{u.line.soliton} 
for the K-KP equation \eqref{KKPeqn.scal}. 
Further investigation of rigorous aspects of a stability proof 
is beyond scope of the present work.

\section*{Acknowledgements}
APM and MLG warmly acknowledge the financial support from the \textit{Junta de Andaluc\'ia} research group FQM-201. SCA is supported by an NSERC Discovery grant.

\end{document}